\documentclass{aip-cp}

\usepackage[numbers]{natbib}
\usepackage{rotating}
\usepackage{graphicx}

\begin{document}

\title{Double Beta Decay to the Excited States: Review}

\author[aff1]{A.S. Barabash\corref{cor1}}

\affil[aff1]{National Research Centre "Kurchatov Institute", Institute of Theoretical and Experimental Physics, B. Cheremushkinskaya 25, 117218 Moscow, Russia}
\corresp[cor1]{barabash@itep.ru}

\maketitle

\begin{abstract}
A brief review of double beta decay to the excited states of daughter nuclei is given. Results of the most sensitive experiments are presented. 
\end{abstract}

\section{INTRODUCTION}
The $\beta\beta$ decay can proceed through transitions to the ground state as well as to various 
excited states of the daughter nucleus. Studies of the latter transitions allow supplementary information about $\beta\beta$ decay. The first experimental study of $\beta\beta$  decay to the excited state has been done by E. Fiorini in 1977 \cite{FIO77}. It was just a by-product of his main experiment 
with $^{76}$Ge (search for transition to 0${^+}$ ground state). First special experimental work to investigate the $\beta\beta$ decay to the excited states has been done in 1982 \cite{BEL82}. In 1989 it was shown that using low-background facilities utilizing High Purity Germanium (HPGe) detectors, the $2\nu\beta\beta$ decay to the 0$^+_1$ level in the daughter nucleus may be detected for such nuclei as $^{100}$Mo, $^{96}$Zr and $^{150}$Nd \cite{BAR90}. Soon after double beta decay of $^{100}$Mo to the 0$^+$ excited 
state at 1130.32 keV in $^{100}$Ru was observed \cite{BAR95}. Then this result was confirmed in a few independent experiments using different detectors and methods \cite{BAR99,DEB01,HOR06,ARN07,KID09,BEL10,ARN14}. In 2004 for the first time this transition has been detected in $^{150}$Nd \cite{BAR04} and then this result has been confirmed in 2014 \cite{KID14}. In addition during the last 15-20 years new limits for many 
nuclei and different modes of decay to the excited states were established (see reviews \cite{BAR07,BAR10,BAR11,LEH15}). Present motivations to do this search are the following: 

I. $2\nu\beta\beta$ decay:

1) nuclear spectroscopy (to know decay schemes of nuclei); 

2) help in solving Nuclear Matrix Elements problem;

3) help in solving $g_A$ problem; 

4) testing of some new ideas (such as the "bosonic" component of the neutrino \cite{DOL05,BAR07a}, for example). 

II. $0\nu\beta\beta$ decay:
 
1) $0\nu\beta\beta$ (0$^+_{g.s.}$ - 0$^+_1$) decay; in this case one has a very nice signature for the decay and hence high sensitivity to neutrino mass could be reached;

2) help in distinguishing between the various $0\nu\beta\beta$ mechanisms; 

3) high sensitivity to the effective Majorana neutrino mass could be reached in the case of the ECEC (0$\nu$) transition if resonance conditions are realized (see \cite{BER83,SUJ04,BAR07b}).

\section{DOUBLE BETA DECAY TO THE EXCITED STATES}

\subsection{$2\nu\beta\beta$ transition to the 2$^+_1$ excited state}

The $2\nu\beta\beta$ decay to the 2$^+_1$ excited state is strongly suppressed and practically inaccessible to detect at present time. However, for a few nuclei ($^{96}Zr$, $^{100}$Mo, $^{130}$Te) there are some "optimistic" predictions for half-lives ($T_{1/2}$ $\sim$ 10$^{21}$-10$^{23}$ y) and there is a chance to detect such decays in the next generation of the double beta decay experiments.
It should also be noted that in the framework of the scheme with "bosonic" neutrinos, the probability of this transition is predicted to increase by approximately an order of magnitude \cite{BAR07a}.
The best present limits are shown in Table 1. 

\begin{table}[ht]
\label{Table1}
\caption{Best present limits on $2\nu\beta\beta$ decay to the 2$^+_1$ excited state (limits at 90\% C.L.). E$_{2\beta}$ is energy of 0$^+$ - 2$^+_1$ transition.}
\vspace{0.5cm}
\begin{tabular}{cccccc}
\hline
Isotope & E$_{2\beta}$, keV & $T_{1/2}$, y & Theory \cite{RAD07} & Theory \cite{PIR15} \\

\hline
$^{48}$Ca & 3279.4 & $>1.8\cdot10^{20}$ \cite{BAK02} & $1.7\cdot10^{24}$ & - \\
$^{150}$Nd & 3037.4 & $>2.2\cdot10^{20}$ \cite{BAR09} & - & $7.2\cdot10^{24}$ \cite{HIR95} \\
$^{96}$Zr & 2577.6 & $>7.9\cdot10^{19}$ \cite{BAR96} & $2.3\cdot10^{25}$ & $(1.1-1.4)\cdot10^{21}$ \cite{SCH98} \\
$^{100}$Mo & 2494.9 & $>2.5\cdot10^{21}$ \cite{ARN14} & $1.2\cdot10^{25}$ & $2\cdot10^{22}$ - 10$^{23}$ \\
$^{82}$Se & 2221.4 & $>1.0\cdot10^{22}$ \cite{BEE15} & $1.7\cdot10^{27}$ & $(1.0-2.4)\cdot10^{24}$ \cite{SCH98} \\
$^{130}$Te & 1991.7 & $>2.8\cdot10^{21}$ \cite{BEL87} & $6.9\cdot10^{26}$ & $(4.2-9.1)\cdot10^{23}$ \\
$^{124}$Sn & 1689.9 & $>9.1\cdot10^{20}$ \cite{BAR08} & - & $(5.3-6.4)\cdot10^{24}$ \\
$^{136}$Xe & 1639.3 & $>4.6\cdot10^{23}$ \cite{ASA16} & $3.9\cdot10^{26}$ & $1.6\cdot10^{25}$ - $4.8\cdot10^{26}$ \\
$^{116}$Cd & 1519.9 & $>2.3\cdot10^{21}$ \cite{PIE94} & $3.4\cdot10^{26}$ & $(2.5-5.2)\cdot10^{24}$ \\
$^{76}$Ge & 1479.9 & $>1.6\cdot10^{23}$ \cite{AGO15} & $5.75\cdot10^{28}$ & $(2.4-4.3)\cdot10^{26}$ \cite{SCH98} \\
\hline
\end{tabular}
\end{table}

\subsection{$2\nu\beta\beta$ transition to the 0$^+_1$ excited state}

For these transitions the best results and limits are presented in Table 2. For $^{100}$Mo and $^{150}$Nd world average values are presented. Table 3 presents all existing positive results for $2\nu\beta\beta$ decay of $^{100}$Mo to the first 0$^+$ excited state of $^{100}$Ru (1130.32 keV). The half-life averaged over all experiments is given in the bottom row. For $^{150}$Nd this transition has been registered in two independent experiments (\cite{BAR04,BAR09} and \cite{KID14}). 
It should be pointed out that in both cases ($^{100}$Mo and $^{150}$Nd) nuclear matrix element for transition to the ground state is approximately 20\% higher then for transition to the excited state and this fact requires an explanation. 
Next most promising candidates are $^{96}$Zr and $^{82}$Se. And tacking into account recent progress in investigation of $^{136}$Xe \cite{ASA16} and $^{76}$Ge \cite{AGO15} one can predict that this transition could be detected in these nuclei too. 

\begin{table}[ht]
\label{Table2}
\caption{Best present results and limits on $2\nu\beta\beta$ decay to the 0$^+_1$ excited state (limits at 90\% C.L.). E$_{2\beta}$ is energy of 0$^+$ - 0$^+_1$ transition.}
\vspace{0.5cm}
\begin{tabular}{cccccc}
\hline
Isotope & E$_{2\beta}$, keV & $T_{1/2}$, y & Theory \cite{PIR15} & Theory \cite{STO96,AUN96,TOI97} \\

\hline
$^{150}$Nd & 2630.9 & $= 1.2^{+0.3}_{-0.2}\cdot10^{20}$ \cite{BAR15} & - & - \\
$^{96}$Zr & 2207.7 & $>3.1\cdot10^{20}$ \cite{FIN15} & - & $(2.4-3.8)\cdot10^{21}$ \\
$^{100}$Mo & 1904.1 & $= 6.7^{+0.5}_{-0.4}\cdot10^{20}$ \cite{BAR15}  & $8.1\cdot10^{21}$ - $4.1\cdot10^{22}$ & $2.1\cdot10^{21}$; $1.6\cdot10^{21}$ \cite{HIR95a} \\
$^{82}$Se & 1510.3 & $>3.4\cdot10^{22}$ \cite{BEE15} & - & $(1.5-3.3)\cdot10^{21}$ \\
$^{48}$Ca & 1265.7 & $>1.5\cdot10^{20}$ \cite{BAK02} & - & - \\
$^{116}$Cd & 1056.6 & $>2.0\cdot10^{21}$ \cite{PIE94} & $(1.6-3.3)\cdot10^{24}$ & $1.1\cdot10^{21}$ - $1.1\cdot10^{22}$ \\
$^{76}$Ge & 916.7 & $>3.7\cdot10^{23}$ \cite{AGO15} & - & $4.5\cdot10^{21}$ - $3.1\cdot10^{23}$ \\
$^{136}$Xe & 878.8 & $>8.3\cdot10^{23}$ \cite{ASA16} & $(1.3-8.9)\cdot10^{23}$ & $(2.5-6.3)\cdot10^{21}$ \\
$^{130}$Te & 734.0 & $>1.3\cdot10^{23}$ \cite{AND12} & $(7.2-16)\cdot10^{23}$ & - \\
$^{124}$Sn & 635.4 & $>1.2\cdot10^{21}$ \cite{BAR08} & $(0.82-1)\cdot10^{25}$ & - \\

\hline
\end{tabular}
\end{table}

\begin{table}[ht]
\label{Table3}
\caption{Present "positive" results on $2\nu\beta\beta$ decay
of $^{100}$Mo  to the first 0$^+$ excited state of $^{100}$Ru (1130.32 keV). N is number of detected events. S/B is the signal-to-background ratio.}
\vspace{0.5cm}
\begin{tabular}{cccc}
\hline
$T_{1/2}$, y & N & S/B & Ref., year \\  
\hline
$6.1^{+1.8}_{-1.1}\cdot 10^{20}$ & 133 & 1/7 & 
\cite{BAR95}, 1995 \\
$[9.3^{+2.8}_{-1.7}(stat) \pm 1.4(syst)]\cdot 
10^{20}$ & 153 & 1/4 & \cite{BAR99}, 1999 \\
$[5.9^{+1.7}_{-1.1}(stat) \pm 0.6(syst)]\cdot 10^{20}$ & 19.5 & $\sim 8$ & \cite{DEB01}, 2001 \\ 
$[5.7^{+1.3}_{-0.9}(stat) \pm 0.8(syst)]\cdot 10^{20}$ & 37.5 & $\sim 3$ & \cite{ARN07}, 2007 \\
$[5.5^{+1.2}_{-0.8}(stat) \pm 0.3(syst)]\cdot 10^{20}$ & 35.5 & $\sim 8$ & \cite{KID09}, 2009 \\ 
$[6.9^{+1.0}_{-0.8}(stat) \pm 0.7(syst)]\cdot 10^{20}$ &  597 & $\sim 1/10$ & \cite{BEL10}, 2010 \\   
$[7.5 \pm 0.6(stat) \pm 0.6(syst)]\cdot 10^{20}$ &  239 & 2 & \cite{ARN14}, 2014 \\   
\hline
{\bf Average value:} $\bf 6.7^{+0.5}_{-0.4}\cdot 10^{20}$ \cite{BAR15} & & \\

\hline
\end{tabular}
\end{table}

\subsection{$0\nu\beta\beta$ transition to the 2$^+_1$ excited state}

The $0\nu\beta\beta$ (0$^+$ - 2$^+_1$) decay had long time been accepted to be possible because of the contribution of right-handed currents and is not sensitive to the neutrino mass contribution. 
However, in Ref. \cite{TOM00} it was demonstrated that the relative sensitivities of (0$^+$ - 2$^+_1$) decays to the neutrino mass ($\langle m_{\nu}\rangle$) and the right-handed current ($\langle \eta\rangle$) are comparable to 
those of $0\nu\beta\beta$ decay to the ground state. At the same time, the (0$^+$ - 2$^+_1$) decay is more sensitive to $\langle \lambda\rangle$. The best present experimental limits are shown in Table 4.

\begin{table}[ht]
\label{Table4}
\caption{Best present limits on $0\nu\beta\beta$ decay to the 2$^+_1$ excited state (90\% C.L.). E$_{2\beta}$ is energy of 0$^+$ - 2$^+_1$ transition.}
\vspace{0.5cm}
\begin{tabular}{cccccc}
\hline
Isotope & E$_{2\beta}$, keV & $T_{1/2}$, y & Theory \cite{TOM00} & Theory \cite{TOM00} \\
&  &  & $\langle m_{\nu}\rangle$ = 1 eV & $\langle \lambda\rangle$ = 10$^{-6}$  \\
\hline
$^{48}$Ca & 3279.4 & $>1.0\cdot10^{21}$ \cite{BAR89} & - & - \\
$^{150}$Nd & 3037.4 & $>2.4\cdot10^{21}$ \cite{ARG09} & - & - \\
$^{96}$Zr & 2577.6 & $>9.1\cdot10^{20}$ \cite{ARN10} & - & - \\
$^{100}$Mo & 2494.9 & $>1.6\cdot10^{23}$ \cite{ARN07} & $6.8\cdot10^{30}$ & $2.1\cdot10^{27}$ \\
$^{82}$Se & 2221.4 & $>1.0\cdot10^{22}$ \cite{BEE15} & - & - \\
$^{130}$Te & 1991.7 & $>1.4\cdot10^{23}$ \cite{ARN03} & - & - \\
$^{124}$Sn & 1689.9 & $>9.1\cdot10^{20}$ \cite{BAR08} & - & - \\
$^{136}$Xe & 1639.3 & $>2.6\cdot10^{25}$ \cite{ASA16} & - & - \\
$^{116}$Cd & 1519.9 & $>6.2\cdot10^{22}$ \cite{DAN16} & - & - \\
$^{76}$Ge & 1479.9 & $>8.2\cdot10^{23}$ \cite{MAI94} & $8.2\cdot10^{31}$ & $6.5\cdot10^{29}$ \\

\hline
\end{tabular}
\end{table}

\begin{table}[ht]
\label{Table2}
\caption{Best present limits on $0\nu\beta\beta$ decay to the 0$^+_1$ excited state (90\% C.L.). E$_{2\beta}$ is energy of 0$^+$ - 0$^+_1$ transition. Theoretical predictions for $\langle m_{\nu} \rangle$ = 1 eV are given. }
\vspace{0.5cm}
\begin{tabular}{cccccc}
\hline
Isotope & E$_{2\beta}$, keV & $T_{1/2}$, y & Theory \cite{KOT12} & Theory \cite{HYV16} \\

\hline
$^{150}$Nd & 2630.9 & $>2.4\cdot10^{20}$ \cite{ARG09} & $1.8\cdot10^{25}$ & - \\
$^{96}$Zr & 2207.7 & $>3.1\cdot10^{20}$ \cite{FIN15} & $8.8\cdot10^{27}$ & $(0.9-1.7)\cdot10^{24}$ \\
$^{100}$Mo & 1904.1 & $>8.9\cdot10^{22}$ \cite{ARN07}  & $2.8\cdot10^{25}$ &  $(6.44-12.5)\cdot10^{25}$ \\
$^{82}$Se & 1510.3 & $>3.4\cdot10^{22}$ \cite{BEE15} & $1.2\cdot10^{26}$ & $(3.79-7.6)\cdot10^{25}$; $1.29\cdot10^{27}$ \cite{MEN09}\\
$^{48}$Ca & 1265.7 & $>1.5\cdot10^{20}$ \cite{BAK02} & $2.3\cdot10^{25}$ & $7.35\cdot10^{25}$ \cite{MEN09} \\
$^{116}$Cd & 1056.6 & $>6.3\cdot10^{22}$ \cite{DAN16} & $(2.5-2.7)\cdot10^{26}$ & $(5.48-12.2)\cdot10^{25}$ \\
$^{76}$Ge & 916.7 & $>1.3\cdot10^{22}$ \cite{MOR88}& $1.4\cdot10^{26}$ & $(4.2-8.28)\cdot10^{25}$; $2.38\cdot10^{26}$ \cite{MEN09} \\
$^{136}$Xe & 878.8 & $>2.4\cdot10^{25}$ \cite{ASA16} & $(5.8-6.4)\cdot10^{25}$ & $(3.7-7.5)\cdot10^{24}$; $5\cdot10^{26}$ \cite{MEN09} \\
$^{130}$Te & 734.0 & $>9.4\cdot10^{23}$ \cite{AND12} & $(3.7-4.4)\cdot10^{25}$ & $(0.53-1.07)\cdot10^{25}$; $6.12\cdot10^{27}$ \cite{MEN09} \\
$^{124}$Sn & 635.4 & $>1.2\cdot10^{21}$ \cite{BAR08} & $1.0\cdot10^{26}$ & $(0.91-1.92)\cdot10^{25}$; $5.82\cdot10^{26}$ \cite{MEN09} \\
\hline
\end{tabular}
\end{table}

\subsection{$0\nu\beta\beta$ transition to the 0$^+_1$ excited state}

The $0\nu\beta\beta$ transition to the 0$^+$ excited states of the daughter nuclei provides a clear cut signature. In addition to two electrons with a fixed total energy, there are two photons, 
whose energies are strictly fixed as well. In a hypothetical experiment detecting all decay products with high efficiency and high energy and spatial resolution, the background can be reduced to nearly zero. It is possible this idea will be used in future experiments featuring a large mass of the isotope under study. In Ref. \cite{SIM02} it was mentioned that detection of this transition will give us the additional possibility to distinguish the $0\nu\beta\beta$ mechanisms. The best present limits are presented in Table 5.

\section{ECEC TO THE EXCITED STATES}

\subsection{ECEC(2$\nu$) transition to the 0$^+_1$ excited state}

In 1994 it was mentioned that there is a possibility to detect ECEC process to the excited 0$^+_1$ state of daughter nuclei and corresponding experiments have been proposed \cite{BAR94}.  Prediction on $T_{1/2}$ for these process is $\sim 10^{22}-10^{23}$ yr for the most promising candidates. During last 20 years  many measurements were done, but this decay still has not been detected.  The best present experimental limits are shown in Table 6. From practical point of view most promising candidates are isotopes with relatively high isotopic abundance ($^{96}$Ru, $^{106}$Cd and $^{112}$Sn).  

\begin{table}[ht]
\label{Table4}
\caption{Best present limits on ECEC(2$\nu$) to the 0$^+_1$ excited state (90\% C.L.). $^{*)}$Estimation from geochemical experiment \cite{MES01}.}
\vspace{0.5cm}
\begin{tabular}{cccccc}
\hline
Isotope & Abundance, \% & Q$_{ECEC(0^+_1)}$, keV & $T_{1/2}$, y & $T_{1/2}$, y; (estimation)  \\
\hline

$^{106}$Cd & 1.25 & 1641.4 & $>1\cdot10^{21}$ \cite{BEL16} & 
$\sim 5\cdot10^{22}$  \\
$^{96}$Ru & 5.54 & 1566.4 & $>2.5\cdot10^{20}$ \cite{BEL13} & $\sim 10^{23}$   \\
$^{78}$Kr & 0.35 & 1349.2 & $>7.5\cdot10^{21}$ \cite{GAV13} & 
$\sim 10^{24}$    \\
$^{124}$Xe & 0.09 & 1199.5 & -  & $\sim 10^{23}$  \\
$^{130}$Ba & 0.106 & 830.2 & $>1.5\cdot10^{21*)}$ \cite{MES01} & 
$\sim 5\cdot10^{23}$      \\
$^{136}$Ce & 0.185 & 799.5 & $>1.6\cdot10^{18}$ \cite{BEL14} & 
$\sim 5\cdot10^{23}$  \\
$^{112}$Sn & 0.97 & 695.4 & $>1.6\cdot10^{21}$ \cite{BAR11a} & $\sim 10^{24}$  \\

\hline
\end{tabular}
\end{table}

\subsection{ECEC(0$\nu$) transition to the excited states}

In Ref. \cite{WIN55} it was first noted that in the case of ECEC(0$\nu$) transition a resonance condition could exist for transition to a "right energy" excited level of the daughter nucleus (when decay energy is close to zero). In 1982 the same idea was proposed for transition to the ground state \cite{VOL82}. In 1983 this possibility was discussed for the 
$^{112}$Sn-$^{112}$Cd (0$^+$; 1871 keV) transition \cite{BER83}. In 2004 the idea was reanalyzed in Ref. \cite{SUJ04}. The possible enhancement of the transition rate was estimated as $\sim$ 10$^6$-10$^8$ \cite{BER83,SUJ04,KRI11,ELI11}. It means that this process starts to be competitive with $0\nu\beta\beta$ decay for the sensitivity to neutrino mass. There are several candidate for such resonance transition, to the ground ($^{152}$Gd, $^{164}$Eu and $^{180}$W) and to the excited ($^{74}$Se, $^{78}$Kr, $^{96}$Ru, $^{106}$Cd, $^{112}$Sn, $^{124}$Xe, $^{130}$Ba, $^{136}$Ce, $^{144}$Sm, $^{156}$Dy, $^{162}$Er, $^{168}$Yb, $^{184}$Os and $^{190}$Pt) sates of daughter nuclei (see 
\cite{KRI11}, for example). The precision needed to realize resonance condition is well below 1 keV. To select the best candidate from the above list one has to know the atomic mass difference with an accuracy better then 1 keV. Such measurements have been done for all mentioned above isotopes. But only in a few cases resonance conditions were found. Unfortunately in all these cases ($^{106}$Cd \cite{GON11} and $^{156}$Dy \cite{ELI12}) there is additional suppression of the decay probability due to "not optimum" quantum numbers of the corresponding excited states (for example, 2$^+$, 2$^-$, 1$^-$,...) and "not optimum" orbits of atomic electrons involved in the process (for example, LL, NM, KM,...). As a result most optimistic prediction for $T_{1/2}$ is on the level $\sim$ 10$^{27}$-10$^{30}$ yr only (for $\langle m_{\nu}\rangle$ = 1 eV) \cite{KOT14,ROD12}. The best present experimental limits for this type of decay are on the level $\sim$ 10$^{21}$ y ($^{112}$Sn \cite{BAR11a}, $^{106}$Cd \cite{BEL16} and $^{78}$Kr \cite{GAV13}).  
But I would like to stress that there is unsatisfactory situation with information about high energy excited states in many nuclei. As a result in same cases there is no reliable information about quantum numbers of the states and their energy (and some levels are just unknown!). So, there is a chance that "promising" candidates for the resonance transition can be found in the future.



\end{document}